\documentclass[prl,twocolumn,showpacs,floatfix]{revtex4}

\usepackage{amssymb}
\usepackage{amsmath}
\usepackage{amsfonts}
\usepackage{graphicx}

\begin{document}

\title{One-spin quantum logic gates from exchange interactions and a
global magnetic field}
\author{Lian-Ao Wu, Daniel A. Lidar}
\affiliation{Chemical Physics Theory Group, and Center for Quantum
  Information and Quantum Control, Chemistry Department, University of Toronto,
80 St. George St., Toronto, Ontario M5S 3H6, Canada}
\author{Mark Friesen}
\affiliation{Department of Physics, University of Wisconsin, Madison, Wisconsin 53706, USA}

\begin{abstract}
It has been widely assumed that one-qubit gates in spin-based quantum
computers suffer from severe technical
difficulties. We show that
one-qubit gates can in fact be generated using only modest and presently
feasible technological requirements. Our solution uses only global magnetic
fields and controllable Heisenberg exchange interactions, thus circumventing
the need for single-spin addressing. 
\end{abstract}

\maketitle

The Heisenberg exchange interaction features prominently in several of the
promising solid state proposals for quantum computation (QC) \cite{Burkard:99,Hu:01a,Levy:01a,Friesen:02,Kane:98,Vrijen:00,Skinner:02}. For this reason extensive studies have been made of
the problem of constructing a universal set of quantum logic gates \cite{Nielsen:book} that employs this interaction. The difficulty is that as long
as each qubit is represented by a single spin it appears necessary to
supplement the Heisenberg interaction with single-spin addressing, thus
requiring either strongly localized magnetic fields \cite{LidarThywissen:04}
or $g$-factor engineering \cite{Kato:03}, both of which are technically
highly demanding. This problem can be alleviated by encoding a logical qubit
into the state of at least three spins, in which case the Heisenberg
interaction by itself can be shown to be universal \cite{Bacon:99a,Kempe:00,DiVincenzo:00a,Hellberg:03}, or by using a combination
of Heisenberg interactions with inhomogeneous, delocalized magnetic fields,
and an encoding into at least two spins \cite{LidarWu:01,Levy:01,Benjamin:01}. Another option is to supplement the Heisenberg interaction with certain
two-spin measurements \cite{WuLidar:02a}. These alternatives to single-spin
addressing each come with space and time overhead \cite%
{Bacon:99a,Kempe:00,DiVincenzo:00a,Hellberg:03,LidarWu:01,Levy:01,Benjamin:01}, or a probabilistic protocol that needs to be repeated several times until
it converges \cite{WuLidar:02a}.

Here we show that in fact it is not necessary to introduce any encoding
overhead when using the Heisenberg interaction in conjunction with
inhomogeneous, delocalized, time-dependent magnetic fields. Furthermore,
unlike previous proposals we do not require spin-resonance techniques \cite{Levy:01} (which involve the difficult demand of modulating the interaction
strength at high frequency), and do not need the recoupling assumption \cite{LidarWu:01} that the exchange coupling can be made much greater than the
difference in Zeeman energies. These considerations overall amount to a
significant simplification compared to various aspects of previously
proposed methods \cite{Bacon:99a,Kempe:00,DiVincenzo:00a,Hellberg:03,LidarWu:01,Levy:01,Benjamin:01}. Instead, one can generate the single-spin gates needed for universality by
pulsing the Heisenberg interaction together with global magnetic fields,
under relaxed controllability assumptions.

\begin{figure*}[tbp]
  \includegraphics[width=16cm]{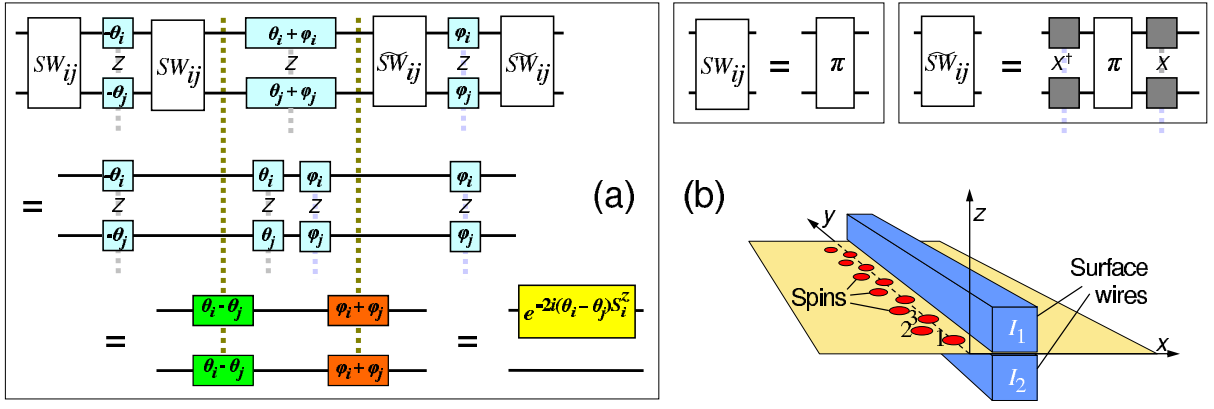}
  \vspace{.3cm}
\caption{
{\bf (a)} Circuit for a one-spin gate.
Time proceeds from left to right. The Heisenberg interaction couples spins $i$ and $j$, carried by the long horizontal lines. Large rectangular boxes
with $\protect\pi$, also denoted by $SW_{ij}$ for ``swap'', represent
$U_{ij}(\pi)$. The action of $\widetilde{SW}_{ij}$ is also a ``swap'', as
explained in the text. Small boxes with angles $\protect\theta _{i},
\protect\theta _{j}$ or $\protect\phi _{i},\protect\phi _{j}$ inside denote
a global magnetic field inducing evolution according to Eq.~(\protect\ref
{eq:V}), and small boxes with $-\protect\theta _{i},-\protect\theta _{j}$
(or $-\protect\phi _{i},-\protect\phi _{j}$) denote $V^{\protect\alpha }(
\protect\theta _{i},\protect\theta _{j})^{\dagger }$ [or $V^{\protect\alpha 
}(\protect\phi _{i},\protect\phi _{j})^{\dagger }$]. Consider an arbitrary single spin
rotation $\exp (-2i\protect\theta S_{i}^{z})$, where the rotation angle $
\protect\theta \equiv \protect\theta _{i}-\protect\theta _{j}$ is
constructed from global rectangular pulses of amplitude $A$ and duration $T$. Neighboring spins experience different amplitudes: $A_{i}=Aa_{i}$, where
the $a_{i}$ are device constants. Typically, $A$ will be limited by physical
considerations, as in the device of (b), so that $T$ controls the pulse
area. Thus, $T=\protect\theta /A(a_{i}-a_{j})$, and the rotation angles
appearing in the circuit are given by $-\protect\theta _{i}=-Aa_{i}T$.
The parameters $\protect\phi _{i}$ and $\protect\phi _{j}$ are defined
through $\protect\theta _{i}-\protect\theta _{j}=\protect\phi _{i}+\protect
\phi _{j}$, so that $\protect\phi _{k}=Aa_{k}T^{\prime }$, where $T^{\prime
}=\frac{a_{i}-a_{j}}{a_{i}+a_{j}}T$. The magnetic field direction $\protect
\alpha $ for each gate is indicated by $X$ or $Z$ between the boxes. Dark
boxes with $\protect\alpha =X$ or $Z$ between them denote $V^{\protect\alpha 
}(\protect\theta ,\protect\theta +\protect\pi )$ with $\protect\theta $
arbitrary, while dark boxes with $X^{\dagger }$ or $Z^{\dagger }$ denote $V^{
\protect\alpha }(\protect\theta ,\protect\theta +\protect\pi )^{\dagger }$.
The circuit yields the gate $e^{-i2(\protect\theta _{i}-\protect\theta 
_{j})S_{i}^{z}}\otimes I_{j}$ ($I$ is the identity operation). Vertical
light dotted lines serve as a reminder that the global magnetic field acts on all
spins, not just spins $i$ and $j$. However, our pulse sequence is
constructed in such a way that the action on all other spins cancels out at
the end of the sequence. This follows from its symmetry and the fact that
the Heisenberg interaction only couples spins $i$ and $j$. The circuit uses 
$11$ elementary steps, four of which are Heisenberg interactions, and seven
of which are global magnetic field pulses. 
{\bf (b)} Zig-zag spin arrangement, with symmetrically positioned wires. The
two current orientations, $I_1=I_2$ and $I_1=-I_2$, enable the complete
control of the global field gradients in the $x$ and $z$ directions.}
\label{fig:pulses}
\end{figure*}

\textit{Interactions}.--- Let us now define the basic interactions involved
in our method. We consider $N$ spins interacting via a controllable
Heisenberg exchange interaction $J_{ij}(t)\mathbf{S}_{i}\cdot \mathbf{S}_{j}$, that generates a unitary evolution 
\begin{equation}
U_{ij}(\xi _{ij})=\exp (-i\xi _{ij}\mathbf{S}_{i}\cdot \mathbf{S}_{j}).
\label{eq:U-Heis}
\end{equation}
Here $\xi _{ij}=\frac{1}{\hbar }\int_{\tau }^{\tau ^{\prime }}J_{ij}(t)dt$
is a fully controllable rotation angle, $\mathbf{S}%
_{i}=(S_{i}^{x},S_{i}^{y},S_{i}^{z})$ and $S_{i}^{\alpha }=\frac{1}{2}\sigma
_{i}^{\alpha }$ ($\alpha =x,y,z$) are the spin-$1/2$ matrices acting on
qubit $i$. In addition we assume the presence of a \emph{global} (i.e.,
delocalized) magnetic field that can be switched on/off fast on the
timescale of spin decoherence. Recent theoretical estimates \cite{deSousa:03}
of dephasing times for electron spins in GaAs quantum dots are on the order
of $50\mu \mathrm{s}$, while experimental data \cite{Tyryshkin:03} for
isotopically pure $^{28}$Si:P at $7$K indicates $T_{2}\gtrsim 60$~ms. We
assume that the field is either spatially inhomogeneous, or there is an
inherent $g$-factor inhomogeneity, or both. In either case, when the
magnetic field is switched on it generates the unitary transformation 
\begin{equation}
V^{\alpha }(\theta _{i},\theta _{j})=\prod_{k=1}^{N}e^{-i\theta
_{k}S_{k}^{\alpha }},  \label{eq:V}
\end{equation}
via the Zeeman effect, where $\theta _{k}=\frac{g_{k}\mu _{B}}{2\hbar }\int_{\tilde{\tau}}^{\tilde{\tau}^{\prime }}B_{k}^{\alpha }(t)dt$. We have
explicitly introduced the indices $i,j$ in Eq.~(\ref{eq:V}) to denote the
pair of spins that is coupled by the Heisenberg interaction in order to
generate a controllable rotation of spin $i$ (see Fig.~\ref{fig:pulses}(a)).
Other notation that we use is $\mu _{B}=9.27\times 10^{-24}\mathrm{{J}{T}^{-1}}$ for the Bohr magneton, and $B_{i}^{\alpha }$ for the value of the
global magnetic field $\alpha $-component at spin $i$. We emphasize that the
only way in which we assume the Zeeman angles $\theta _{i}$ to be
controllable is through the time-dependence of the magnetic field, \emph{which is the same for all spins}: $B_{i}(t)=f(t)B_{i}$, where $f(t)$ does
not depend on the spin index $i$ and $B_{i}$ does not depend on time. The
inhomogeneity assumption means that $\theta _{i}\neq \pm \theta _{i+1}$.

\textit{Circuits}.--- With these definitions we are ready to present our
analytically derived pulse sequence that generates arbitrary single-spin
operations. It is given as a quantum circuit in
Fig.~\ref{fig:pulses}(a). It is simple to verify the validity of the
circuit by direct matrix multiplication. The circuit of Fig.~\ref{fig:pulses}(a) shows that an arbitrary one-spin $S^{z}$-operation can be
generated in terms of controllable Heisenberg interactions and a global
time-dependent magnetic field in $11$ elementary steps. Below, we show how
to implement the required magnetic field using a simple spin architecture,
with two nearby wires (see Fig.~\ref{fig:pulses}(b)). We can also generate an
arbitrary $S^{x}$ operation, simply by switching the current direction in
one of the wires. Since these operations generate $SU(2)$ on an arbitrary
spin, and the Heisenberg interaction can generate entanglement, \emph{we
have a universal set of quantum logic gates}. An explicit circuit for
generating a controlled-phase (CP) gate can be given as follows:
\[
e^{-i \pi S_{i}^{z}S_{j}^{z}} = U_{ij}(\pi/2) V^{z}(\theta,\theta +\pi )^{\dagger }U_{ij}(\pi/2) V^{z}(\theta,\theta +\pi ),
\]
which requires four elementary steps, and is equivalent to CP$=\mathrm{diag}(1,1,1,-1)$ up to single-qubit
corrections (a circuit that generates CP up to an overall
phase takes $11$ steps). This procedure is an application of the 
familiar refocusing technique of NMR
(see, e.g., \cite{Nielsen:book}), wherein the
terms $S_{i}^{x}S_{j}^{x}+S_{i}^{y}S_{j}^{y}$ that are also part of the
Heisenberg interaction, are cancelled.

Intuitively, the role of the Heisenberg pulses in Fig.~\ref{fig:pulses}(a) is
to \emph{swap} the states of the spins they act on: $U_{ij}(\pi
)e^{-i(\theta _{i}S_{i}^{z}+\theta _{j}S_{j}^{z})}U_{ij}(-\pi )=e^{-i(\theta
_{i}S_{j}^{z}+\theta _{j}S_{i}^{z})}$. For this reason we denote $U_{ij}(\pi
)$ by $SW_{ij}$ in Fig.~\ref{fig:pulses}(a). The operator represented by $%
\widetilde{SW}_{ij}=V^{x}(\theta ,\theta +\pi )^{\dagger }U_{ij}(\pi
)V^{x}(\theta ,\theta +\pi )$ in 
Fig.~\ref{fig:pulses}(a) also acts as a kind of swap
operator, in that, it is easy to show that it transforms 
$\widetilde{SW}_{ij}e^{-i(\theta_{i}S_{i}^{z}
+\theta _{j}S_{j}^{z})}\widetilde{SW}_{ij}=ie^{i(\theta _{i}S_{j}^{z}
+\theta _{j}S_{i}^{z})}$.
These swaps are the fundamental operations that allow one to address single
spins in the presence of a global magnetic field with a gradient, since they
allow a spin to be exposed to different values of the field without needing
to localize the field to a particular spin site. More specifically, the
swaps allow us to modify the relative phase acquired by the two target spins
under the action of the global magnetic field. By appropriately modifying
the signs of these phases, as shown in Fig.~\ref{fig:pulses}(a), we can
completely cancel the evolution of one spin while generating a desired
unitary transformation on the other spin. In spirit, the method is again
similar to refocusing.

\textit{Parallelism}.--- Our scheme is parallelizable in the following
sense: \emph{the }same \emph{gate operation can be applied to all pairs of
spins that are not Heisenberg-coupled}. This follows simply by virtue of our
assumed ability to simultaneously switch the Heisenberg interaction in
different locations, coupled with the \emph{global} coupling to the magnetic
field. The limitation to applying the same operation imposes some
constraints in terms of the parallelism required for fault-tolerant quantum
error correction; how the fault tolerance threshold \cite{Steane:02} is
affected under such conditions is an issue we will consider in a future
publication. Note, however, that an identical gate operation applied to all
qubits is a common feature of quantum algorithms; e.g., a global Hadamard
transform is used to create a superposition over all computational basis
states starting from the initial state $|00\cdots 0\rangle $. Our analysis
shows that such a global Hadamard transform would take only eight elementary
steps for appropriate qubit geometries.

\textit{Other exchange models}.--- Our results are not restricted to the
Heisenberg interaction. Consider, e.g., the XY model (which describes a
range of QC proposals, e.g., quantum dots in microwave cavities \cite{Imamoglu:99}), where the exchange Hamiltonian takes the form $%
\sum_{i<j}J_{ij}(S_{i}^{x}S_{j}^{x}+S_{i}^{y}S_{j}^{y})$. The circuit for
generating $e^{-i\theta _{i}S_{i}^{z}}$ is identical to the one shown in 
Fig.~\ref{fig:pulses}(a), with the Heisenberg Hamiltonian replaced by the XY\ one. One
has $e^{i2\theta _{i}S_{i}^{x}}=V^{x}(\theta _{i},\theta _{j})^{\dagger
}e^{i\pi S_{i}^{z}}V^{x}(\theta _{i},\theta _{j})e^{i\pi S_{i}^{z}}$ and
similarly for $S_{i}^{y}$. The following steps will then generate a
controlled-phase gate:
\begin{eqnarray*}
  e^{i\frac{\pi }{4}(S_{i}^{y}-S_{j}^{y})}
  e^{-itJ(S_{i}^{x}S_{j}^{x}+S_{i}^{y}S_{j}^{y})} e^{i\pi
    S_{i}^{x}} \times \\
  e^{-itJ(S_{i}^{x}S_{j}^{x}+S_{i}^{y}S_{j}^{y})} e^{i\pi
  S_{i}^{x}} e^{-i\frac{\pi}{4}(S_{i}^{y}-S_{j}^{y})} = e^{-i2tJZ_{1}Z_{2}}.
\end{eqnarray*}
The corresponding circuit takes $32$ steps, and can certainly be optimized.
Another important exchange model is the Heisenberg interaction with
time-independent spin-orbit corrections included. The ``dressed qubits'' method of Ref. \cite{WuLidar:03} is
then compatible with the circuits of Fig.~\ref{fig:pulses}(a).

\textit{Implementation}.--- We now address the technical feasibility of our
method. Besides the (conventional) requirement of precise pulse timings, it
may appear that the most serious technical challenge in our scheme (as can
be seen from Fig.~\ref{fig:pulses}) is that we need a very large magnetic
field gradient (or $g$-factor difference) because of the $\pi $ difference
in terms such as $V^{x}(\theta ,\theta +\pi )$. Assuming for simplicity a
rectangular $B$-field pulse of duration $\Delta T$, this condition
translates into $(B_{i+1}-B_{i})\Delta T=18~\mathrm{mT\,ns}$, where we took
a uniform $g$-factor of $2$. To achieve a typical gate time \cite{Friesen:02}
of $\Delta T\simeq 10$~ns, we require a magnetic field increment of $%
B_{i+1}-B_{i}=1.8$~mT. A uniform field gradient of this magnitude, extending
across a scalable device of $\sim 10^{4}$ spins, is not physically
reasonable: eddy current heating will occur as the large field is switched.
Instead, we now suggest a simple scheme to implement the needed gradients,
using on-chip patterned wires. Such microfabricated conducting wires are
extensively used in the atom optics and BEC physics communities, where the
technology required to manufacture them has reached maturity \cite{Reichel:02}.

Two superconducting wires are positioned near an array of spins, as shown in
Fig.~\ref{fig:pulses}(b). The spins are arranged in a zig-zag pattern, with half
the spins located at $x=0$ and the other half at $x=-100$~nm (this, and
other numbers used below, should be taken as suggestive estimates; our
scheme is general and does not depend on these specific values). Note that
for clarity we show only a single row of spins. Our method is fully
compatible with a 2D architecture, where the zig-zag pattern is repeated
parallel to the first row. Inspection of Fig.~\ref{fig:pulses}(a) shows that if
the exchange interaction is never switched on then the global magnetic field
pulses exactly cancel. Thus, to address a specific row of spins one simply
does not turn on the exchange interaction in any other row (additional wires
may have to be placed between rows of spins as the number of rows increases,
due to the $1/r$ fall-off of the field strength). A tunable exchange
coupling is established between consecutive spins in the row to be
addressed. The wires of cross-section $200\times 200$~nm are centered on
either side of the spin plane at $x=200$~nm and $z=\pm 100$~nm, with a thin
insulating layer inbetween. To avoid resistive heating, the wires should be
operated below their superconducting critical currents. For nano-patterned
aluminum wires, we can expect \cite{Anthore:03} critical current densities
of order $2.2\times 10^{10}$~A/m$^{2}$. Thus, our wires can safely carry
currents of size $I_{1,2}=0.7$~mA. Magnetic fields arising from such
currents range well below the superconducting critical fields \cite{Anthore:03}. 

We consider two particular current configurations: (a) $I_{1}=I_{2}=0.7$~mA,
and (b) $I_{1}=-I_{2}=0.7$~mA. Due to symmetry, we find that $B^{x}=0$ in
case (a), with a discrete magnetic field gradient of $%
|B_{i+1}^{z}-B_{i}^{z}|=0.28$~mT. For case (b), we obtain $B^{z}=0$ and $%
|B_{i+1}^{x}-B_{i}^{x}|=0.28$~mT. Maximum pulse durations are therefore
given by $\Delta T\simeq 64$~ns. The two current configurations above yield
arbitrary single-spin $S^{x}$ and $S^{z}$ operations. Observing that we need
at most $3\times 7=21$ magnetic field pulses to realize an arbitrary
single-spin operation (using the Euler angles method \cite{Nielsen:book}),
we deduce a total gate time of about $1.3$~$\mu $s. Considering that we have
not attempted serious optimization and have used limitations imposed by
current experiments, this number compares quite favorably with the fault
tolerance requirement \cite{Steane:02} of $10^{-3}-10^{-5}$ times the
recently estimated decoherence time for impurity-bound spins, $T_{2}\gtrsim
60$~ms \cite{Tyryshkin:03}. However, we note that since our scheme is
somewhat limited in the ability to parallelize single-spin operations (the
angles $\theta _{i},\theta _{j}$ of two spins $i,j$ involved in an exchange
interaction in a given row cannot be independently controlled), and since
the exchange interaction enables effective coupling of nearest-neighbor
spins only, the effective threshold applicable to our case is likely to be
lower \cite{Gottesman:99a} than the current optimal estimates \cite{Steane:02}.

We can also use fault tolerance limits to back out engineering
specifications for the wire fabrication. Small inaccuracies in wire
positions will result in systematic rotation errors. In this case, fault
tolerance suggests an \emph{amplitude} accuracy \cite{Gottesman:comment} of about $\sqrt{10^{-4}}$ for single qubit rotations, composed of at most 21 magnetic
field pulses. In the worst case scenario pulse errors accumulate coherently,
and we will require an accuracy of $10^{-2}/21$ per pulse. For the wiring
scheme of Fig.~\ref{fig:pulses}(b), this accuracy level corresponds to
fabrication precision of about 1~\AA , or roughly one atomic monolayer of
aluminum. This should be considered a lower bound; in practice, after
optimization and considering a more realistic error model, we expect the
required fabrication precision to be on the order of several atomic
monolayers, which is technologically feasible.

We now show that other potential technical obstacles with the proposed 
architecture can be overcome.  We have estimated the eddy current heating of electrons in 
the 2DEG, due to the rapid switching of magnetic fields of 0.6~mT. We find 
that only extremely short switching times, $\Delta T < 2$~ps, are capable 
of generating enough heat to disturb low-temperature quantum dot 
experiments \cite{Heis:note}.  Further, our target switching times of tens of ns 
cannot excite the low-temperature superconducting gaps of aluminum or 
niobium.  Considering a conventional gating scheme \cite{Friesen:02}, we 
estimate that transient voltage fluctuations in the gate electrodes caused 
by field switching are 7-9 orders of magnitude weaker than normal gate 
pulses.  Additionally, we estimate the $B$~fields produced by transient 
currents during the pulsed charging of gate electrodes.  For typical pulses 
of 0.1~V, with 0.1~ns rise times, we obtain $B$~field fluctuations three 
orders of magnitude smaller than those from the wires, and extremely brief 
compared to gate times.  

Finally, we note there already exists an extensive literature on the
important issues of spin readout \cite{Hanson:03,Manassen:00-Berman:01-Ciorga:01-Fujisawa:02-Xiao:03} and accurate
control of the exchange coupling \cite{Burkard:99,Hu:01a,Levy:01a,Friesen:02,Kane:98,Vrijen:00,Skinner:02}, needed
to implement our scheme.

\textit{Conclusions}.--- We have shown how to construct a universal set of
quantum logic gates using the Heisenberg exchange interaction between
nearest neighbour spins and \emph{global}, inhomogeneous magnetic fields.
Our scheme bears some similarity to the global addressing schemes developed
in the context of quantum cellular automata \cite{Benjamin:02}, a
connection we plan to explore in a future publication. Optimization of our
pulse sequences appears feasible and will be undertaken in a future study.
However, our estimates indicate that even before optimization the method we
have proposed is feasible using current technology, and is compatible with
the time-scale requirements of fault tolerant quantum error correction. We
believe that this alternative to previous methods for QC in
exchange-Hamiltonian systems offers a significant simplification and a
drastic reduction in design constraints. We thus hope to have brought QC
with spins a step closer to experimental feasibility.

We are grateful to Prof.~Mark Eriksson and Dr.~Alexandre Zagoskin for useful
discussions and thank the DARPA-QuIST program, the NSF-QuBIC program, and
ARO/ARDA for financial support.


\begin{thebibliography}{10}

\bibitem{Burkard:99}
{G. Burkard, D. Loss and D.P. DiVincenzo}, Phys. Rev. B {\bf 59},  2070
  (1999).

\bibitem{Hu:01a}
{X. Hu, R. de Sousa and S. Das Sarma}, Phys. Rev. Lett. {\bf 86},  918  (2001);
{X. Hu and S. Das Sarma}, Phys. Rev. A {\bf 64},  042312  (2001).

\bibitem{Levy:01a}
{J. Levy}, Phys. Rev. A {\bf 64},  052306  (2001).

\bibitem{Kane:98}
{B.E. Kane}, Nature {\bf 393},  133  (1998).

\bibitem{Vrijen:00}
{R. Vrijen {\it et al}.}, Phys. Rev. A {\bf 62},  012306  (2000).

\bibitem{Friesen:02}
{M. Friesen {\it et al}.}, Phys. Rev. B {\bf 67},  121301(R)  (2003).

\bibitem{Skinner:02}
{A.J. Skinner, M.E. Davenport, B.E. Kane}, Phys. Rev. Lett. {\bf 90},  087901
  (2003).

\bibitem{Nielsen:book}
{M.A. Nielsen and I.L. Chuang}, {\em {Quantum Computation and Quantum
  Information}} ({Cambridge University Press}, Cambridge, UK, 2000).

\bibitem{LidarThywissen:04}
{D.A. Lidar, J.H. Thywissen}, J. Appl. Phys. (in press), eprint cond-mat/0310352.

\bibitem{Kato:03}
{Y. Kato {\it et al}.}, Science {\bf 299},  1201  (2003).

\bibitem{Bacon:99a}
{D. Bacon, J. Kempe, D.A. Lidar and K.B. Whaley}, Phys. Rev. Lett. {\bf 85},
  1758  (2000).

\bibitem{Kempe:00}
{J. Kempe, D. Bacon, D.A. Lidar, and K.B. Whaley}, Phys. Rev. A {\bf 63},
  042307  (2001).

\bibitem{DiVincenzo:00a}
{D.P. DiVincenzo {\it et al}.}, Nature {\bf
  408},  339  (2000).

\bibitem{Hellberg:03}
C. Hellberg, eprint quant-ph/0304150.

\bibitem{LidarWu:01}
{D.A. Lidar and L.-A. Wu}, Phys. Rev. Lett. {\bf 88},  017905  (2002).

\bibitem{Levy:01}
{J. Levy}, Phys. Rev. Lett. {\bf 89},  147902  (2002).

\bibitem{Benjamin:01}
{S.C. Benjamin}, Phys. Rev. A {\bf 64},  054303  (2001).

\bibitem{WuLidar:02a}
{L.-A. Wu, and D.A. Lidar}, Phys. Rev. A {\bf 67},  050303  (2003).

\bibitem{deSousa:03}
{R. de Sousa and S. Das Sarma}, Phys. Rev. B {\bf 67},  033301  (2003).

\bibitem{Tyryshkin:03}
{A.M. Tyryshkin, S.A. Lyon, A.V. Astashkin, A.M. Raitsimring}, eprint
  cond-mat/0303006.

\bibitem{Steane:02}
{A.M. Steane}, \pra {\bf 68}, 042322 (2003).

\bibitem{Imamoglu:99}
{A. Imamo$\bar{\rm g}$lu {\it et al}.}, Phys. Rev. Lett. {\bf 83},  4204  (1999).

\bibitem{WuLidar:03}
{L.-A. Wu and D.A. Lidar}, Phys. Rev. Lett. {\bf 91},  097904  (2003).

\bibitem{Reichel:02}
{J. Reichel}, Appl. Phys. B {\bf 75},  469  (2002).

\bibitem{Anthore:03}
{A. Anthore, H. Pothier and D. Esteve}, Phys. Rev. Lett. {\bf 90}, 127001,  
  (2003).

\bibitem{Gottesman:99a}
D. Gottesman, J. Mod. Optics {\bf 47},  333  (2000).

\bibitem{Gottesman:comment}
D. Gottesman, private communication. We use the conservative threshold
  estimate of $10^{-4}$ for the probability of error per logic gate, yielding
  an error amplitude of $10^{-2}$.

\bibitem{Heis:note}
We have gauged excessive heat levels from low-temperature experiments on
  coupled point contact and quantum dots, where a maximum of $10$~pW was
  dissipated over a volume $\sim 10^{5}~\mathrm{nm}^{3}$. See
  \protect\cite{Hanson:03}.

\bibitem{Hanson:03}
{R. Hanson {\it et al}.}, Phys. Rev. Lett. {\bf 91},  196802  (2003).

\bibitem{Manassen:00-Berman:01-Ciorga:01-Fujisawa:02-Xiao:03}
{Y. Manassen, I. Mukhopadhyay, and N. Ramesh Rao}, Phys. Rev. B {\bf 61},
  16223  (2000);
{G.P. Berman, G.W. Brown, M.E. Hawley, V.I. Tsifrinovich}, Phys. Rev. Lett.
  {\bf 87},  097902  (2001);
{M. Ciorga {\it et al}.}, Physica E {\bf 11},  35  (2001);
{T. Fujisawa {\it et al}.}, Nature {\bf 419},  278  (2002);
{M. Xiao, I. Martin, and H.W. Jiang}, Phys. Rev. Lett. {\bf 91},  078301
(2003).

\bibitem{Benjamin:02}
{S.C. Benjamin}, Phys. Rev. Lett. {\bf 88},  017904  (2002).

\end{thebibliography}

\end{document}